# Ti$_3$C$_2$T$_x$ MXene Enabled All-Optical Nonlinear Activation Function for On-Chip Photonic Deep Neural Networks

Adir Hazan[1], Barak Ratzker[2], Danzhen Zhang[3], Aviad Katiyi[1], Nachum Frage[2], Maxim Sokol[4], Yury Gogotsi[3] and Alina Karabchevsky[1,*]

[1] School of Electrical and Computer Engineering, Electro-Optics and Photonics Engineering Department, Ben-Gurion University of the Negev, Beer-Sheva 8410501, Israel.
[2] Materials Engineering Department, Ben-Gurion University of the Negev, Beer-Sheva 8410501, Israel.
[3] Materials Science and Engineering Department, and A.J. Drexel Nanomaterials Institute, Drexel University, Philadelphia 19104, USA.
[4] Materials Science and Engineering Department, Tel Aviv University, Ramat Aviv 6997801, Israel.
[*] E-mail: alinak@bgu.ac.il

## Abstract

Neural networks are one of the first major milestones in developing artificial intelligence systems. The utilisation of integrated photonics in neural networks offers a promising alternative approach to microelectronic and hybrid optical-electronic implementations due to improvements in computational speed and low energy consumption in machine-learning tasks. However, at present, most of the neural network hardware systems are still electronic-based due to a lack of optical realisation of the nonlinear activation function. Here, we experimentally demonstrate two novel approaches for implementing an all-optical neural nonlinear activation function based on utilising unique light-matter interactions in 2D Ti$_3$C$_2$T$_x$ (MXene) in the infrared (IR) range in two configurations: 1) a saturable absorber made of MXene thin film, and 2) a silicon waveguide with MXene flakes overlayer. These configurations may serve as nonlinear units in photonic neural networks, while their nonlinear transfer function can be flexibly designed to optimise the performance of different neuromorphic tasks, depending on the operating wavelength. The proposed configurations are reconfigurable and can therefore be adjusted for various applications without the need to modify the physical structure. We confirm the capability and feasibility of the obtained results in machine-learning applications via an Modified National Institute of Standards and Technology (MNIST) handwritten digit classifications task, with near 99% accuracy. Our developed concept for an all-optical neuron is expected to constitute a major step towards the realization of all-optically implemented deep neural networks.

**Introduction**

Artificial neural networks (NNs) are computational network models which are inspired by the way the signals are processed in the brain. NNs have been implemented in numerous integrated photonics applications. These include the optical response prediction of subwavelength nanophotonic devices[1], neuromorphic computing[2], obtaining the inverse design for a given optical response[3–6], single-pixel cameras that capture coded projections of a scene with a single photodetector and computationally recover them[7], and others. The utilisation of integrated photonics in NNs[8] offers a promising alternative approach to microelectronic and hybrid optical-electronic implementations, owing to the improvement in computational speed and power efficiency in machine-learning tasks. While implementing NNs optically, the challenge remains in implementing the nonlinear activation function - which is currently still being fulfilled electronically[9–14] and costing a great deal in time and power consuming operations. To address this challenge, here we utilise the integrated photonics platform and nonlinear properties of MXene to demonstrate a novel concept of the NN activation function implemented all-optically[15–17]. Integrated photonics provides a stable, compact and robust platform to implement complex photonic circuits[18], and therefore is a compelling technology for the realisation of photonic NN with a time delay on the order of picoseconds.

Unlike conventional nonlinear activation function mechanisms, which are based on devices that convert an optical signal into an electrical signal, then by applying the nonlinearity convert it back into an optical signal - resulting in slower computational speed and noise - we utilise the optical effects of novel 2D material MXene.

The integration of the nonlinear optical unit in a photonic circuit remains an open challenge, along with strengthening modulation and increasing operational speed, to increase the effectiveness at both the device and system levels. To address this challenge, we propose mechanisms of implementing NN activation function based on the electrooptic effect, which does not require electrical and temperature controls. Recently, theoretical work on an all-optical nonlinear activation function based on a saturable absorber was proposed in refs[15,19] but has never been proved experimentally. Here, we demonstrate the nonlinear activation function on a chip. The concept of an all-optically implemented nonlinear activation function on a chip that we propose is expected to be the most energy-efficient and to operate at a five-fold higher processing speed as compared to existing conventional NNs implemented electronically and can therefore compete with conventional von Neumann computer architecture.

MXenes comprise a large class of 2D transition metal carbides and nitrides such as $Ti_3C_2T_x$ ($T_x$ represents surface terminations such as -O, -OH and -F)[20,21]. They exhibit unique light-matter interactions such as the nonlinear effect[22] of saturable absorption[23] on one hand, and plasmonic properties on the other hand[24,25]. Furthermore, integrating MXene in photonic circuits is extremely useful for the all-optical nonlinear activation function in NN.

Here, we demonstrate the optical neuron nonlinear activation function based on nanophotonic structures. Specifically, we use: 1) a saturable absorber made of $Ti_3C_2T_x$ MXene thin film, and 2) a nanophotonic waveguide covered with MXene flakes. We test them experimentally and develop an NN-based emulator to analyse the results. The nonlinear activation function with executed MNIST handwritten digit classification task reported here is of 99.1% accuracy.

**Results**

**Nanophotonic neural network mechanism of operation**

The concept of the fundamental building block for a fully integrated deep structured learning, as illustrated in Fig. 1a, represents an entire layer in deep-NN (DNN) using a MXene overlayer on a chip to introduce nonlinear properties in the network. The optical signals carrying encoded information that propagates via a linear combination of the inputs arriving from the preceding building block (i.e., a layer) pass through the nonlinear activation function. This function eventually serves as an input to the next building block unit. In general, the photonic building blocks can optically implement arbitrary DNN and are defined as a function $f: \mathbb{R}^n \to \mathbb{R}^m$, where $n$ and $m$ ($n, m \geq 1$) are the number of neurons in the input and output layers, respectively. As shown schematically in Fig. 1b, a well-known deep-learning architecture consists of input and output layers, with at least one hidden layer. Generally, a feedforward network contains interconnected neurons that do not form a cycle (neurons connect only to different neurons in successive layers). The operating principle of each neuron in the network can be divided into the following two optical interfaces: 1) the linear operation unit involves weighting and summation, and 2) the nonlinearity unit defines the neuron output by generating a new signal given an input.

An optical interference unit (OIU) can be realised using various integrated photonics architectures to implement matrices multiplication for weighting and summation. The physical implementation can be classified as optical modes realisation such as linear operation nanophotonics circuits[10,26–29], or multiwavelength realisation such as parallel weighting of optical carrier signals generated from wavelength-division multiplexing using microring resonators weight banks[30–34]. Those have been demonstrated previously during different computational tasks. For multiple weighted $W_{ij}^{(l)}$ input signals $n_j^{(l-1)}$ arriving from the output of neurons in the previous layer with the addition of a bias $b_i^{(l)}$, the optical linear interface of the $i^{th}$ neuron $n_i^{(l)}$ in the layer $l^{th}$, is given by a linear operation across all the inputs, i.e., $n_i^{(l)} = b_i^{(l)} + \sum_j W_{ij}^{(l)} n_j^{(l-1)}$. A realistic

OIU building block may be emulated in DNN modelling by considering the analytical form of a whole layer output $X_{OIU}^{(l)}$, as follows: $X_{OIU}^{(l)} = W^{(l)} Y^{(l-1)}$, through the forward-propagation procedure. For instance, weighted input signals can be implemented with a nanophotonic circuit of integrated Mach–Zehnder interferometers, each formed of waveguides and 50:50 directional couplers which are combined with phase shifters[10,26]. In particular, where any unitary transformations can be implemented with conventional optical beamsplitters and phase shifters[35], a rectangular diagonal matrix can be implemented with optical attenuation achieved by Mach–Zehnder modulator. As well, an OIU implementation that relies on free-space diffractive DNN[36–38] has already been shown in the spectral domain[39,40]. When diffraction light interference implements the weighting, a sum of these signals is achieved through combined transmission (or reflection) coefficients at each point on a given transmissive layer that acts as a neuron. However, an OIU alone is insufficient for a photonic device to act as a building block in DNN applications, and some optical nonlinearities must be introduced. In principle, a photonic neuron generating a new optical output signal processing the multiple optical inputs signals $X_{OIU}^{(l)}$, through the nonlinear activation function, $n_i^{(l)} = f_{NL}(X_{OIU}^{(l)})$. Implementations of optical nonlinearity unit (ONLU) fall into two major categories, one based on optical-to-electrical-to-optical (OEO) and the other on all-optical (AO). With the help of light-MXene interaction, the nonlinear activation function can be realised in an AO manner.

**Design and fabrication of MXene-based all-optical nonlinear activation function**

Generally, an artificial neuron includes multiple weighted input signals followed by nonlinear activation. Focusing on the nonlinear neuron unit, we study its all-optical nonlinear activation functions utilising unique light-matter interactions in 2D $Ti_3C_2T_x$ – MXene. Here, we validate the photonic DNN performance by focusing on the shape of the activation functions. To this end, the linear operation is emulated with the nonlinear optical activation function, realised with two independent configurations. For this purpose, we engineered two devices to introduce the all-optical nonlinear activation function. Figure 2a and inset of Fig. 4a show rendered images of studied architectures to demonstrate the concept of an all-optical nonlinear activation function with an MXene overlayer. A non-polarised electromagnetic wave illuminates the in-facet of a multimode rib-waveguide (Fig. 2a). The interaction with the MXene overlayer takes place via evanescent waves. The unpolarised plane electromagnetic wave illuminates the thin film of MXene on a glass substrate, as shown in the inset of Fig. 4a. To study the nonlinear response of fabricated samples, two experimental setups operating in a broad spectral range were constructed. For MXene thin films, a coherent supercontinuum generation laser source was focused on the film, and then the light was collected by an optical spectrum analyser (OSA) via a fibre. For the on-chip configuration, a nanophotonic waveguide covered with MXene flakes was butt-coupled via single-mode fibre, then the light was collected by OSA via a multimode fibre. In addition, the waveguide surfaces were imaged on the camera for inspection, characterisation and alignment.

**Formation of MXene flake-based metasurface on a waveguide**

One possibility for inducing an optical nonlinearity in a photonic integrated circuit is by exploiting a hybrid system consisting of a silicon waveguide with a MXene flake overlayer. Figure 2a shows a rendered image of studied architecture on a chip to demonstrate the concept of an all-optical nonlinear activation function with an MXene overlayer. Non-polarised electromagnetic wave illuminates the in-facet of a multimode rib-waveguide. The interaction with the MXene overlayer takes place via evanescent waves and 1) leads to some losses within the material, and 2) causes the scattering of light in all directions. The waveguide is wide enough to support multiple modes to increase the coupling between the evanescent waves and MXene nano-flakes. This can be achieved with the higher-order modes that have a longer evanescent field extension into the medium and larger field amplitude at the waveguide cladding interface, compared to the fundamental and lower modes. To produce an MXene metasurface (a metasurface is a patterned thin film composed of elements at a subwavelength scale to achieve tailored properties), the first $Ti_3C_2T_x$ MXene was synthesised through selective chemical etching using the LiF + HCl method[41]. Then, to realise an MXene-based metasurface, a diluted MXene suspension in water with 0.01 g/ml was prepared and drop-casted on a waveguide. Figure 2b shows a schematic illustration describing the synthesis process of MXenes from MAX phases and redispersion of the dry product to produce MXene dispersed suspension.

Concurrently, we fabricated a silicon-on-insulator rib waveguide with an MXene flakes overlayer. Figure 3a shows the schematics of an inline experimental setup operating with broadband illumination to test such waveguides. Figure 3b shows transmission spectra collected via multimode fibre from the distal end of the waveguides covered by an MXene flake-based metasurface with well-pronounced spectral signatures: a dip around 1180 nm with the depth of -25 dB and a dip around 1490 nm with the depth of -20 dB. These dips can be associated with plasmonic excitation oriented MXene flakes on the waveguide surface. Plasmonic excitation in MXene arises from a plasmon-induced increase in the ground state absorption at photon energies above the threshold for free carrier oscillations[23]. The dip in transmission spectrum around 1490 nm, shown in Fig. 3b, can be associated with the first overtone excitation of -OH functional group out-of-plane vibrations of MXene[42]. The first principle calculations[43] verifies the fundamental vibration related to this overtone. The dip in transmission around 1180 nm can be associated with the waveguide shifted overtone vibration of MXene metasurface assigned to the $OH/H_2O$ native oxide layer on the waveguide surface, or with plasmonic excitation, because the real part of the permittivity of MXene is negative in this range - as can be seen from the dispersion spectra we measured with ellipsometry (shown in Supplementary Fig. S1). In addition, the nanoscale flakes of MXene can exhibit a nano-antenna effect resulting in resonances appearing in transmission spectra. The shallow peak around 980 nm corresponds to the minor peak shown in Fig. 3b and can be explained by the MXene dispersion as measured with an ellipsometer (see Supplementary Fig. S1).

Figure 3c shows a scanning electron micrograph of the top view on reference waveguides, while Fig. 3d shows a scanning electron micrograph of the top view on waveguides covered by an MXene flake-based metasurface overlayer. To better understand the light-matter interaction, specifically the interaction between the evanescent waves and MXenes overlayer, we numerically explored a unit cell effect made of two MXene nanodiscs atop the silicon, illuminated by the evanescent waves studied rib waveguide. Calculated results show the extinction cross-section curve as in Supplementary Fig. S3a with two peaks. The smaller peak appears at a shorter wavelength (around 1020 nm), while the larger peak appears at a longer wavelength (around 1560 nm). Supplementary Fig. S3b shows the power loss density in a slice through the MXene nanodiscs to assess the intensity dependence in the proposed hybrid system.

**MXene saturable absorber**

A further alternative approach to realise the nonlinear activation function is utilising the optical nonlinearity via the effect of saturable absorption, for which the absorption decreases with an increase in the input light intensity. This could be expressed by the material absorption coefficient at a given wavelength as $\alpha = \frac{\alpha_0}{1+I/I_S}$, where $\alpha_0$ is the linear absorption coefficient, $I$ and $I_S$ are the incident and saturation intensities. Recently, 2D $Ti_3CNT_x$ was found to exhibit nonlinear saturable absorption (or increased transmittance) at higher light fluences[23,44]. In addition, it was shown that the saturation fluence and modulation depth of $Ti_3CNT_x$-MXene depend on the film thickness[23]. To experimentally extract the nonlinear optical response, we fabricated four spray-coated samples of thin films of $Ti_3C_2T_x$ on BK-7 glass, with an increasing thickness between 50 nm and 90 nm. One may expect to observe the saturable absorption property of MXene thin-film via free-space illumination measurement. For this, we used unpolarised light illuminating a 50 nm MXene film on BK-7 substrate and collected via the focusing objective into the multimode optical fibre directly connected to the optical spectrum analyser illustrated in Fig. 4a. The measured transmission spectra of MXene thin-film response to input powers vary from 6% to 80%, as shown in Fig. 4b. The calculated linear optical transmittance of the $Ti_3C_2T_x$ films on a glass substrate as a function of wavelength (Fig. 4c) decreases with an increase in the film thickness, which correlates well with experimental measurements. The model considers the measured refractive index and extinction coefficient for different thicknesses, with surface roughness measured by a profilometer. When the incident intensity at which the absorption coefficient is half of the linear absorption coefficient (i.e., $\alpha = \alpha_0/2$) define as the saturation intensity, which is dependent on the film thickness (Fig. 4d, top). We note that the random roughness of spin-coated films (Fig. 4d, bottom) does not affect the observed nonlinear transmission effect. The modulation depth, as depicted in Fig. 4e and can be determined by the maximum change in saturable absorber for a given wavelength as follows: $\Delta T_{NL}(\lambda) = \frac{T_{NL}(\lambda) - T_L(\lambda)}{T_L(\lambda)}$ , where $T_{NL}$ and $T_L$ are nonlinear and linear transmissions, respectively.

**Operation of the nonlinear activation function.**

As a first step, we experimentally achieved a set of transmission spectrum measurements for each nonlinear activation function mechanism. The transmission spectrum was monitored on an optical spectrum analyser to observe the nonlinear responses by controlling the input optical power. Each set consists of several measurements for various input powers from 6% to 96%. Proceeding further, we obtain transfer functions that represent the instantaneous input and output power amplitudes measured at a specific wavelength (detailed in the Methods section). The wavelength dependence of optical responses allows us to tune the nonlinear activation function shape. These transfer functions represent the device's nonlinear optical responses by output power vs input power relation. We note that these observed nonlinear functions represent a subset of functions achievable by our devices without modifying the structure of the devices. A generic activation function squashes a real input number to a fixed interval specified with unitless (Supplementary Fig. S4). In contrast, realistic optical activation function input and output units are in the context of spectral quantities specified with units of mW/nm. Thus, in this hardware-software co-investigation, they are regarded as the relation of power in and power out. Here, we aim to validate the performance of well-known DNN architectures by focusing on the nonlinear operation shape formed by the studied photonics devices through examining their functionality in a conventional machine learning task. Therefore, the model avoids noises generated by the device or these induced by the OIU (through the summation of the weighted input signals), which could, in a future investigation, be scaled for large-scale NN deployment.

**All-optical neural network emulation**

Going forward, we introduced the obtained nonlinear optical responses as neuron's nonlinear activation function in a conventional machine-learning task: a handwritten digit image to be classified. Figure 5a shows the studied structure of feedforward DNN comprising two hidden layers, each comprising of 100 neurons, and resulting in outputs in the range of from 0 to 9. Thus, the output layer in our network has only ten neurons belonging to one of ten categories, representing digits from 0 to 9. We aim to identify the representing numbers for each input image using DNN, as shown in Fig. 5a; several outputs predicted labels correspond to four input handwritten digit images. Feeding each image to the input layer requires preprocessing each two-dimensional matrix of a handwritten digit image to a high-dimensional vector. Then, the input signals can be encoded in the amplitude of optical pulses when propagating through the photonic integrated circuit. Each layer of the DNN is composed of optical interference and nonlinearity units, which implement optical matrix multiplication and nonlinear operation, respectively. As discussed earlier, the input optical signals are weighted and combined through a mesh of integrated Mach-Zehnder interferometers. However, employing an MXene metasurface overlayer on waveguide or MXene thin film configurations can achieve the nonlinear activation function. In addition, we noted that between two consecutive layers, on each output connection, the nonlinear activation function is placed (e.g. each neuron sums all the weighted inputs from neurons in the proceeding layer and then applies the nonlinear activation function).

By emulating the behaviour of the experimentally implementing nonlinear optical operations of the studied approaches as neuron's nonlinear activation function in photonic DNN, one can effectively evaluate their functionality. To this end, we utilised the Tensorflow platform to emulate the photonic DNN's performance in terms of accuracy and loss compared to those obtained with software-based nonlinear activation functions for the MNIST dataset. In particular, we trained our networks with two sets of nonlinear activation functions: the proposed all-optical devices and software-based commonly used in machine-learning applications. These transform functions are obtained from experimentally measurements at various operating wavelengths for MXene metasurface overlayer on waveguide and MXene thin film configurations, represent power in -to power out relation as shown in Fig. 5b. In contrast, Supplementary Fig. S4 shows the commonly used software-based nonlinear activation function we employed to compare our photonic configurations - the Rectified Linear Unit (ReLU), Softplus, Exponential linear units (ELU), and Mish. To achieve the mathematical model of the nonlinear operations used in photonic neural network emulation, we considered an $f_{NL}$ that models the transfer functions associated with each studied mechanism due to MXene-light interaction (details presented in the Methods section).

During the training and testing process, we considered three separate datasets. While the training dataset randomly broke down into two subsets, 80% and 20% (e.g. 48,000 and 12,000 images, respectively), to trained and validated the model. The testing dataset ensures the model can classify the images without acknowledging the data is beforehand, based on learning about the data features. It is worth mentioning, through the validation process, the weights in the model are not updating based on the loss calculated. Figure 5c shows the network prediction accuracy when the network is trained for 50 epochs, indicating that all the proposed all-optical nonlinear activation functions achieve competitive classification accuracies against the standard software-based nonlinear activation functions on the MNIST dataset. The validation data accuracy and loss verify the training dataset (Supplementary Fig. S5). All the proposed photonic nonlinear operations considering both MXene meta-surface overlayer on waveguide and MXene thin film achieve accuracies between 97.9% to 99.1%.

To better understand the functionality of the proposed nonlinear operation mechanisms as compared to the well-established and commonly used nonlinear activation functions, we emulated their behaviour in different network architecture, utilising a convolutional NN (CNN). The task of the CNN is the same, requiring it to identify the representing numbers for each input image of the MNIST handwritten digits dataset. This class of networks operates under a different approach as compared to the first studied network. As depicted in Fig. 5d, the studied CNN is decomposed into unique layers. The input is a handwritten digit image that is processed by convolutions, pooling, operation of the proposed all-optical nonlinear activation function, a fully connected layer followed by a softmax activation, and last, the output resulting in the range of 0 to 9. Figure 5e represents an accuracy comparison between the proposed all-optical MXene-based and software-based nonlinear activation functions during the training and validation processes. Also, in this case of CNN, one can see that the proposed activation mechanisms provide competitive performance in terms of accuracy and loss as a function of epoch with respect to the well-established and commonly used nonlinear activation functions. After training and validation of the networks, they predicted the output of unlabeled data in the testing dataset. Figure 5f shows the experimental confusion matrices of the MNIST classification, using the transfer functions of MXene thin film (blue) and MXene metasurface overlayer on waveguide (purple) at an operating wavelength of 1550 nm, achieved 98.9% and 97.4% recognition accuracy of the test data set, respectively. Moreover, besides this classification example, there are other applications where such nonlinear activation functions are routinely used for artificial neural network tasks and in a wide variety of types of networks, in addition to the two examples shown here.

Taking into account the availability of dozens of stoichiometric and solid-solution MXenes with a wide range of optical properties and plasmon resonances covering the wavelength from UV to IR, there are attractive prospective for designing photonic devices with MXenes beyond $Ti_3C_2T_x$.

## Conclusions

We have demonstrated an all-optical nonlinear activation function operating in a wide spectral range. We fabricated $Ti_3C_2T_x$ MXene thin films and MXene overlayers on waveguides and compared the optical response in the cases of: 1) evanescent excitation; 2) plane wave illumination. We numerically explored a unit cell effect made of two MXene nanodiscs atop the silicon waveguide. The resulting transmission spectrum dips can be explained as localised surface plasmon excitation at wavelengths of 1020 nm and 1560 nm. Our emulator showed compatible performance of the proposed activation mechanisms based on a MXene metasurface overlayer on the waveguide and a MXene thin film, in terms of accuracy and loss as a function of epoch with respect to the well-established and commonly used nonlinear activation functions in machine-learning tasks. The nonlinear response of the activation function was achieved due to the saturable absorber property of MXene. The all-optically implemented activation function demonstrated in this work may open a new chapter in the development of neural networks for efficient computation.

## Methods

**MXene thin films preparation.** $Ti_3C_2T_x$ was synthesised by the selective etching of $Ti_3AlC_2$ MAX phase powder (325 mesh) with a mixture of HF (48.5-51%, Acros Organics) and HCl (36.5−38%, Fisher Chemical) acids[45]. Typically, 2 mL of HF, 12 mL of HCl, and 6 mL of deionised (DI) water were mixed. After that, 1 g of MAX phase powder was added to the solution and stirred for 24 h at 35 °C. After etching, the reaction product was washed with DI water using the centrifuge at 3500 rpm for 2 min until pH >6. The obtained sediment was dispersed in a 0.5 M LiCl solution. The mixture was shaken for 15 min and then centrifuged at 3500 rpm for 10 min several times until the sediment was delaminated and swelled. The swelled sediment was dispersed in DI water and then centrifuged at 3500 rpm for 10 min. After that, the dark supernatant containing primarily single-layer MXene sheets was collected for spray-coating. Finally, four thin films of ~50 nm to ~90 nm thicknesses $Ti_3C_2T_x$ were spray-coated on a borosilicate glass (BK-7) substrate with a ratio of 4.5 mg/ml DI water.

**MXene thin films characterisation.** To characterise the surface roughness and thickness of the fabricated MXene films, topography measurements were performed with the Stylus profilometer, Veeco Dektak-8.

**Ellipsometric spectrometry.** We obtained the optical parameters of MXene, namely, the refractive index $n$ and extinction coefficient $\kappa$ via spectroscopic ellipsometer. Spectroscopic ellipsometer measurements were performed in the wavelength range: 245-1690 nm. The samples consisted of a BK-7 glass substrate with $Ti_3C_2T_x$ coating of approximate thicknesses of 50 nm, 67 nm, 72 nm, and 91 nm.

**Waveguide fabrication.** The rib waveguides were fabricated as detailed in ref [46], based on a Silicon-On-Insulator (SOI) wafer with silicon Carrier, 2 μm of silica $SiO_2$ and 2 μm of silicon. E-beam resist poly-methyl methacrylate (PMMA) 950k was used together with a line pattern mask via a conventional photolithography process. Once the PMMA resist was developed, aluminum was evaporated to serve as a hard mask with the thickness of 250 nm via an Electron Gun evaporator. Next, we soaked the chip in acetone for four hours (lift-off process) and cleaned the chip with isopropanol. Eventually, the chip was dry-etched with $SF_6$+Ar and $O_2$ to achieve straight lines and 90-degree waveguide walls. The residue of the Al hard mask was removed with a 400K developer.

**Procedure for preparing MXene flakes for waveguide overlayer.** The concept of waveguide overlayer is schematically shown in Fig. 2a and the preparation procedure in Fig. 2b. The $Ti_3AlC_2$ powders were synthesised by mixing titanium carbide (Alfa Aesar, 99.5% 2 microns), aluminum (Alfa Aesar, 99.5%, 325 mesh), and titanium (Alfa Aesar, 99.5%, 325 mesh), powders in a molar ratio of 2:1.1:1, respectively. The powders were mixed in a horizontal rotary mixer at 100 rpm for 24 h and then heated under Ar flow at 1400 °C for 3 h. The heating and cooling rates were set at 5 °C/min. The resulting loosely sintered block was ball milled to powders and passed through a 400 mesh (< 38 μm) sieve. The $Ti_3AlC_2$ powder was etched in a LiF and HCl solution. Initially, 1 g of LiF (Alfa Aesar, 99.5%, 325 mesh) was dissolved in 10 mL of 12 M HCl (Fisher Scientific). Later, 1 g of the $Ti_3AlC_2$ powder was slowly added to the solution and stirred for 24 h at 35 °C and 300 rpm. After etching the slurry was transferred into a 50 ml centrifuge tube and deionised (DI) water was added to fill the remaining volume. It was then centrifuged at 2300 rcf for 2 min and the resulting clear supernatant was discarded. The same washing process was repeated several times until the pH of the solution was ~7, at which point DI water was added to the resulting $Ti_3C_2T_x$ "clay" and the mixture was sonicated under bubbling Ar flow for 1 h. To avoid oxidation, the bath temperature was kept below 20 °C using ice. The solution was then centrifuged for 1 h at 4700 rcf and the supernatant was pipetted off, dried in a drying oven at 120 °C for 12 h and sealed under Ar for storage and future use. To obtain the MXene flakes solution, 0.1 g of dry $Ti_3C_2T_x$ was added to 10 ml DI water and sonicated in an ultrasonic bath for 5 min, resulting in a solution of dispersed $Ti_3C_2T_x$ suspension with a concentration of 0.01 g/ml.

**Scanning electron microscopy.** The surfaces SEM micrographs of blank reference waveguides and metasurface overlayer of MXene on a rib waveguide were examined with a high-resolution scanning electron microscope (FEI Verios 460L).

**Numerical simulation.** We calculated the absorption and extinction cross-section profiles of the nanodisks atop the waveguide numerically. The three-dimensional simulation was carried out using a commercial COMSOL Multiphysics 5.6 software based on the finite element analysis method in wave optics module, as a unit cell with periodic boundary conditions. Mesh was explored to ensure the accuracy of the calculated results. The dielectric constant of the material entirely defines the material optical properties. Therefore, the empirical dielectric functions of the silicon and silica were taken from the Refractive-Index database (https://refractive index.info). In contrast, the dielectric function of MXene (Supplementary Fig. S4) obtaining via the measured refractive index distribution with a spectroscopic ellipsometer. In the simulation, the thickness of the MXene nanodisks was set to 10 nm with a radius of 0.25 µm as evaluated from SEM images of the fabricated MXene flakes.

**Experimental systems for nonlinear transmission measurements.** Two experimental systems were used to measure the nonlinearity in the optical response of the two MXene configurations. Both setups are used for achieving a broadband spectrum of the transmitted light through the proposed configurations using standard optical communication components. All setups were constructed in a cleanroom environment. The energy source for optical computing was generated using a supercontinuum white-light laser source (SuperK EXTREME EXW-12, NKT Photonics), bandwidth from 390 nm to 2400 nm, fibre delivered and collimated with an output power of 5.5 W. The beam was focused on single-mode fibre (P1-1550A-FC, 1460 - 1620 nm, ⌀125 µm cladding, Thorlabs) using an ×10 infinity-corrected imaging microscope objective (RMS10X, with a numerical aperture of NA = 0.25, Olympus).

For silicon WG covered with MXene flakes configuration, inline measurements setup was used with butt-coupled light from a single-mode fibre to the input waveguide facet. The output optical spectra were collected via the conventional silica multimode fibre (MMF 50:125 µm core to cladding respectively) directly into the optical spectrum analyser (AQ6370D, Yokogawa), as shown in Fig. 3a. The fibres were held with 3-axis piezoelectric stages that allow flexibility for precise adjustment of the fibres to the waveguide input and output facets. In addition, the waveguide was imaged (top view) onto the camera (Axiocam, ZEISS) using a microscope (Stemi SV 6, ZEISS) for accurate inspection, characterisation and alignment. Prior to the measurements, the MXene flakes solution was prepared with a ratio of 1 mg/100 ml DI water. Then, a droplet of 2 μl of the solution was dripped atop the nanophotonic rib waveguide using a micropipette and dried up in the cleanroom environment.

MXene thin film characterisation was performed with the transmission setup we built, shown in Fig. 4a. The coherent supercontinuum generation light with constant pump power at the set level was collimated via a protected silver reflective collimator (RC04FC-P01, 450 nm-20 μm, ⌀4 mm beam, Thorlabs), then passed through an iris. The sample was mounted on a fixed stage and illuminated by the unpolarised supercontinuum generation white light, with a spot size of 100 μm onto the MXene nano-film on a BK-7 glass substrate. The transmitted light was collected into the optical spectrum analyser via MMF using an ×4 infinity-corrected imaging microscope objective (RMS4X, with a numerical aperture of NA = 0.1, Olympus). The laser was operated in modulated power mode that generated picosecond pulses with a repetition rate of 78.56 MHz.

**Spectrometry measurements.** To observe the optical responses due to MXene-light interaction, the intensity of the transmitted light when the MXene is present was first measured. As a reference measurement, the spectra without the contribution of MXene were collected. The differential transmission spectra were then plotted (Fig.3b and Fig.4b), which are given by:

$$\Delta T(\lambda) = \frac{|T_{\text{MXene}}(\lambda)|^2}{|T_{\text{Ref}}(\lambda)|^2} \tag{1}$$

In the case of MXene flakes overlayer on a waveguide, $|T_{\text{MXene}}|^2$ is the transmittance when an unpolarised light is coupled to a rib waveguide with a presence of $Ti_3C_2T_x$ on the top surface, whereas $|T_{\text{Ref}}|^2$ is the transmittance spectra collected from a blank reference waveguide. In the case of MXene thin films $|T_{\text{MXene}}|^2$ is the transmittance when the unpolarised light hits the BK-7 substrate, which is covered with $Ti_3C_2T_x$ nano-film, whereas $|T_{\text{Ref}}|^2$ is the transmittance through the glass medium. In each case, ten measurements were carried out to follow the changes in input power.

**Activation function settings.** To obtain a mathematical function that modelled the transfer function of the all-optical activation function to be used in photonic neural network emulation, we fit data points from the experimental results to the total broadband optical transmittance of the devices.

For the MXene-waveguide configuration, we fit quadratic curves due to the nonlinear operation acting on the optical intensity, which is directly related to the electric field amplitude with squaring proportionality. The total transmittance is defined fundamentally by the power losses within the interaction length of the MXene-waveguide. Therefore, the transmittance through the MXene flakes overlayer on a waveguide system is obtained as in ref[47]:

$$T(\lambda) = \left| \sum_{\gamma 1=i,j,m} C_{\gamma 1} \exp(-i\alpha_{\gamma 1} L) \right|^2 \tag{2}$$

where $C_{\gamma 1} = (I_{\gamma 0,\gamma 1} + I_{\gamma 1,\gamma 0})^2 / (4 I_{\gamma 0,\gamma 0} I_{\gamma 1,\gamma 1})$, $L$ is the interaction length, $\alpha_{\gamma 1}$ is an attenuation coefficient of modes in a region covered with MXene flakes, $\gamma 1$ are the guided modes influenced by the MXene, and $\gamma 0$ are the guided modes in pure dielectric waveguide.

For the MXene thin film configuration, we utilise the saturable absorption property of MXene. A saturable absorber material is characterised by the dependence of its absorption on the incident laser intensity. Therefore, at a given wavelength $\lambda$, the transmission through the MXene thin film can be expressed as follows[48]:

$$T(I) = 1 - \Delta T_{NL} \exp\left(\frac{-I}{I_S}\right) - T_{\text{ns}} \tag{3}$$

where $\Delta T_{NL}$ is the modulation depth, $I$ and $I_S$ are the incident and saturation intensities, and $T_{\text{ns}}$ is the initial transmittance of the absorber defined fundamentally by the nonsaturable loss in the material and depends on the design of the saturable absorber.

**Photonic neural network emulation.** We modelled the photonic DNN with an end-to-end open-source platform for machine learning TensorFlow, for handwritten digit classification tasks. The MNIST handwritten digit dataset consists of 60,000 and 10,000 images belonging to training and testing, respectively. Each image is composed of 28x28 pixel resolution associated with one of ten categories representing numbers in the range of 0 to 9. We split the training set into two subsets of 80% (training set) and 20% (validation set) images for trained and validated the model. While the weights are subsequently optimised in the training process using the backpropagation algorithm[49], the validation set was used to validate the network without weights updating. The photonic DNN is trained by feeding data into the input layer, then based on the loss calculated from output prediction (so-called forward propagation), optimising weights using a backpropagation algorithm using a stochastic gradient descent method. The network model used a stochastic gradient descent optimiser with a learning rate of 0.001. After forming two network architectures, we evaluate their performance using typical nonlinear activation functions. Proceeding further, we emulated the photonic NN considering our nonlinear operation based on their transfer functions. By emulating these proposed all-optical nonlinear operations, one can estimate the effect of the all-optical nonlinear activation function on the overall functionality of the NN. The transfer functions represent the device's nonlinear optical responses by output power to input power relationships. In general, software-based nonlinear activation functions are unitless that define a nonlinear output to input relation. Therefore, we considered he obtained transfer functions as power out to power in relation, where the actual values in the context of spectral quantities specified with units of mW/nm (or dBm/nm). Finally, we compared the prediction accuracy and loss of the networks employing our all-optical nonlinear activation functions to those achieved with the commonly used software-based activation functions (as shown in Fig 5., Supplementary Fig. S5 and S6). These results conclude that our proposed all-optical nonlinear operations provide comparative performance as those successfully adopted in the machine learning community.

## Data availability

The data that support the plots within this paper and other findings of this study are available from the corresponding author upon reasonable request.

## Acknowledgements

A.K.* acknowledges the support of the Israel Science foundation ISF grant no. 2598/20. D.Z. and Y.G. work on MXene development was supported by a grant from the US National Science Foundation DMR- 2041050.

## Author contributions

A.K.* and A.H. conceived the idea; A.H. conducted the numerical studies; A.K. constructed the experimental setup; A.H., A.K., A.K* conducted the experimental study. A.H. prepared the initial draft of the manuscript. Y.G. and D.Z. developed MXene thin films on BK-7 glass. M.S., B.R. and N.F. developed Mxene flakes in solution; A.K.* supervised the project. All co-authors discussed the results. A.K.* and A.H. wrote the manuscript with input from all co-authors.

## Competing interests

The authors declare no competing interests.

## Additional information

**Supplementary Information.** A file contains supplementary material is available.

**Correspondence and requests for materials** should be addressed to A.K.*

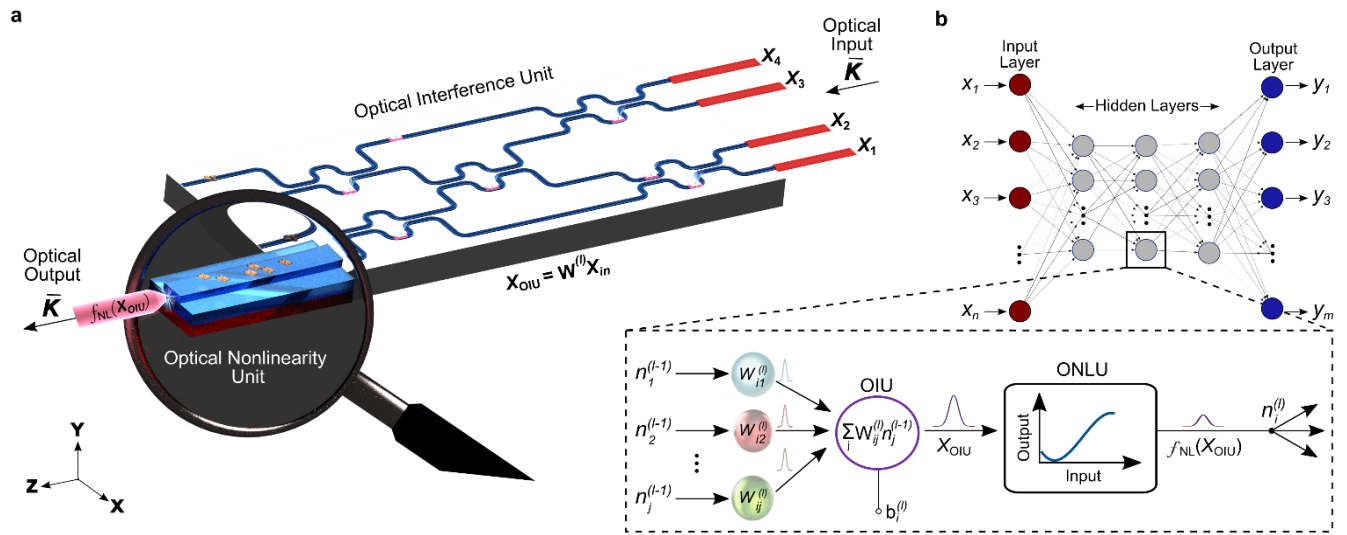

**Figure 1 | Schematic conceptualisation of fully integrated MXene-based DNN building block. a,** The rendered chip shows the implementation of a whole layer in DNN via a photonic circuit (grey rectangle). The input laser (red) encodes the information carried through the chip via waveguides (blue). The chip relies on a mesh of Mach–Zehnder interferometers; each consists of directional couplers (curved sections in the waveguides integrated with phase shifters (red glowing objects) to control the splitting ratio and the differential output phase. The zoom-in illustrates the neuron's nonlinear activation function based on light-MXene interaction, where a rib waveguide is covered with MXene flakes. The arrows indicate the propagation direction of the light. **b,** General architecture of fully connected DNN constituted of an input layer (red circles), several hidden layers (gray circles), and an output layer (blue circles); Inset shows a schematic of one neuron, the input optical signals are weighted and combined, then the nonlinear activation function is applied.

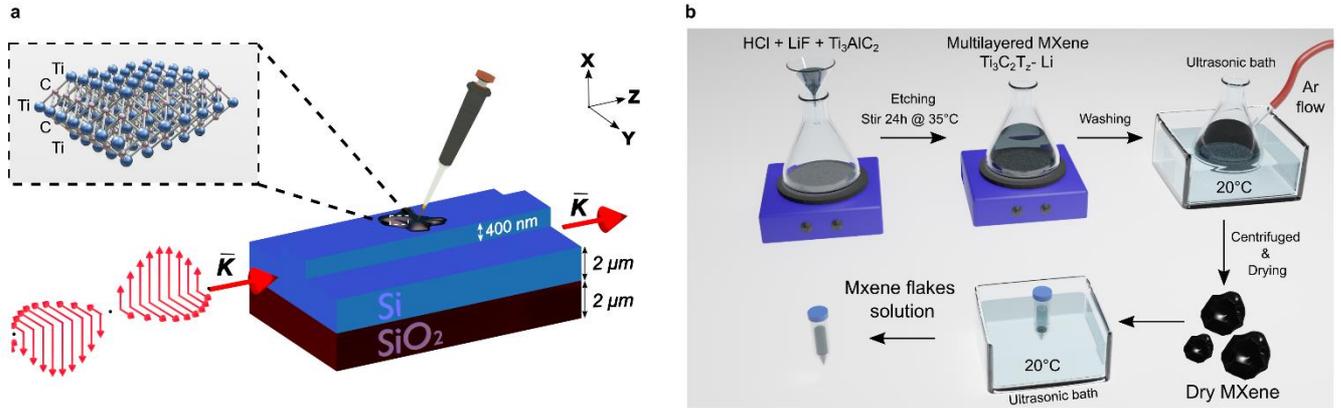

**Figure 2 | MXene flakes production for waveguide overlayer. a,** The studied system's artistic design: incident unpolarised light hits the nanophotonic rib waveguide facet, which is covered with MXene flakes. The inset is showing the crystal structure of a 2D MXene monolayer, $Ti_3C_2T_x$. **b,** Schematic illustration describing the synthesis process of MXene from MAX phases and redispersion of the dry product to produce MXene dispersed suspension.

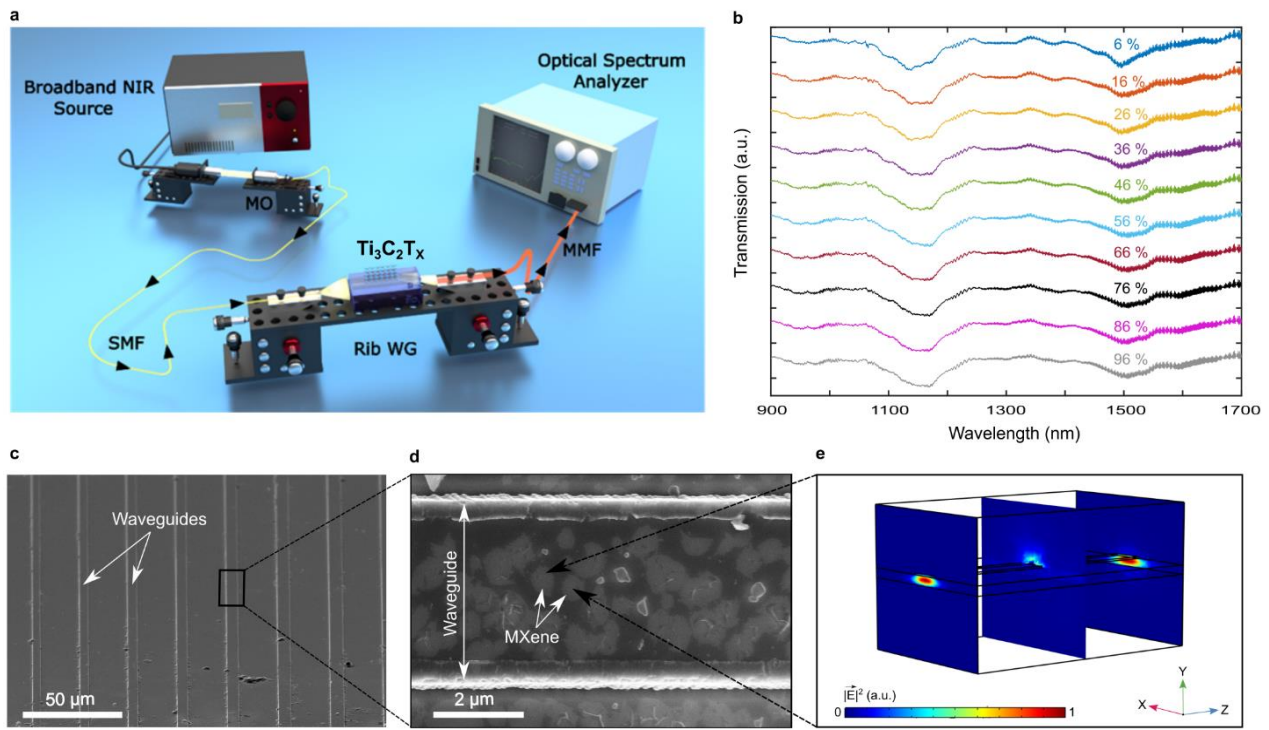

**Figure 3 | Experimental evidence of all-optical nonlinear activation function on a chip. a,** Schematics of the experimental setup with broadband input light coupled to the waveguide and collected from waveguide output facet via optical fibre into a spectrum analyser. **b,** Measured transmission spectra from silicon rib waveguide covered with MXene flakes for input power varying from 6% to 96% (from top to bottom). **c-d,** Top view scanning electron microscope (SEM) images of blank reference waveguides (**c**) and metasurface overlayer of MXene on a rib waveguide (**d**). **e,** Calculated field distribution of optical mode propagating in the waveguide core and interacting with MXene nanodiscs indicated by arrows in subplot **d** with the dispersion of MXene (in Supplementary section).

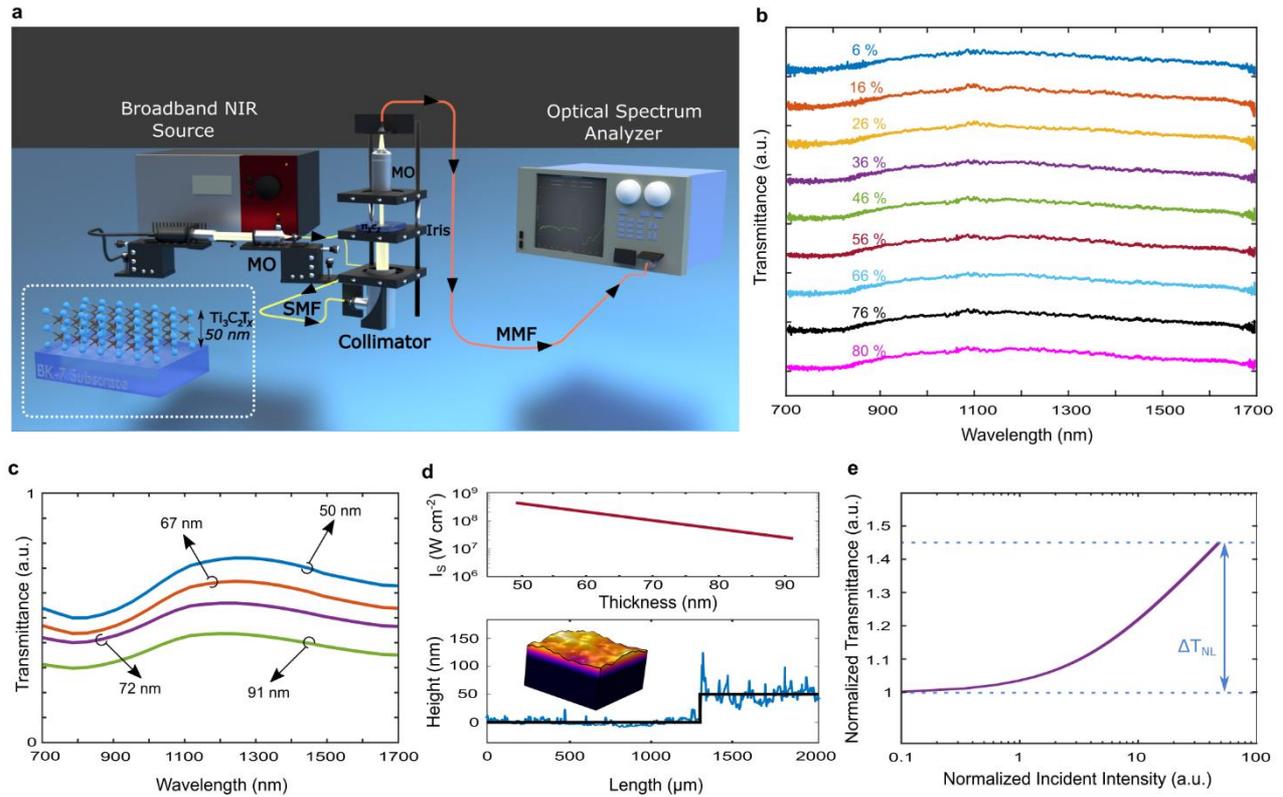

**Figure 4 | Free-space response of MXene thin films to input power. a,** Rendered transmission setup with inset showing a thin film of $Ti_3C_2T_x$ on BK-7 glass; Microscope objective (MO), Single-mode fibre (SMF), Multimode fibre (MMF). **b,** Measured transmission spectra for MXene thin films with 50 nm thickness on BK-7 substrate, for various input power from 6% to 80% (top to bottom). **c,** Calculated transmission spectra for MXene films with thicknesses (top to bottom) of 50 nm, 67 nm, 72 nm, and 91 nm. **d,** (top) The saturation intensity vs MXene thickness at the wavelength of 1550 nm; (bottom) The measured thickness of MXene with profilometer (average marked by the black line) compared to the modelled random roughness of MXene thin film on BK-7 substrate of the depth of 12.7 nm. **e,** Nonlinear transmission of the 50 nm MXene film as a function of input intensity evaluated from the transmission waveguide spectroscopy at the wavelength of 1550 nm.

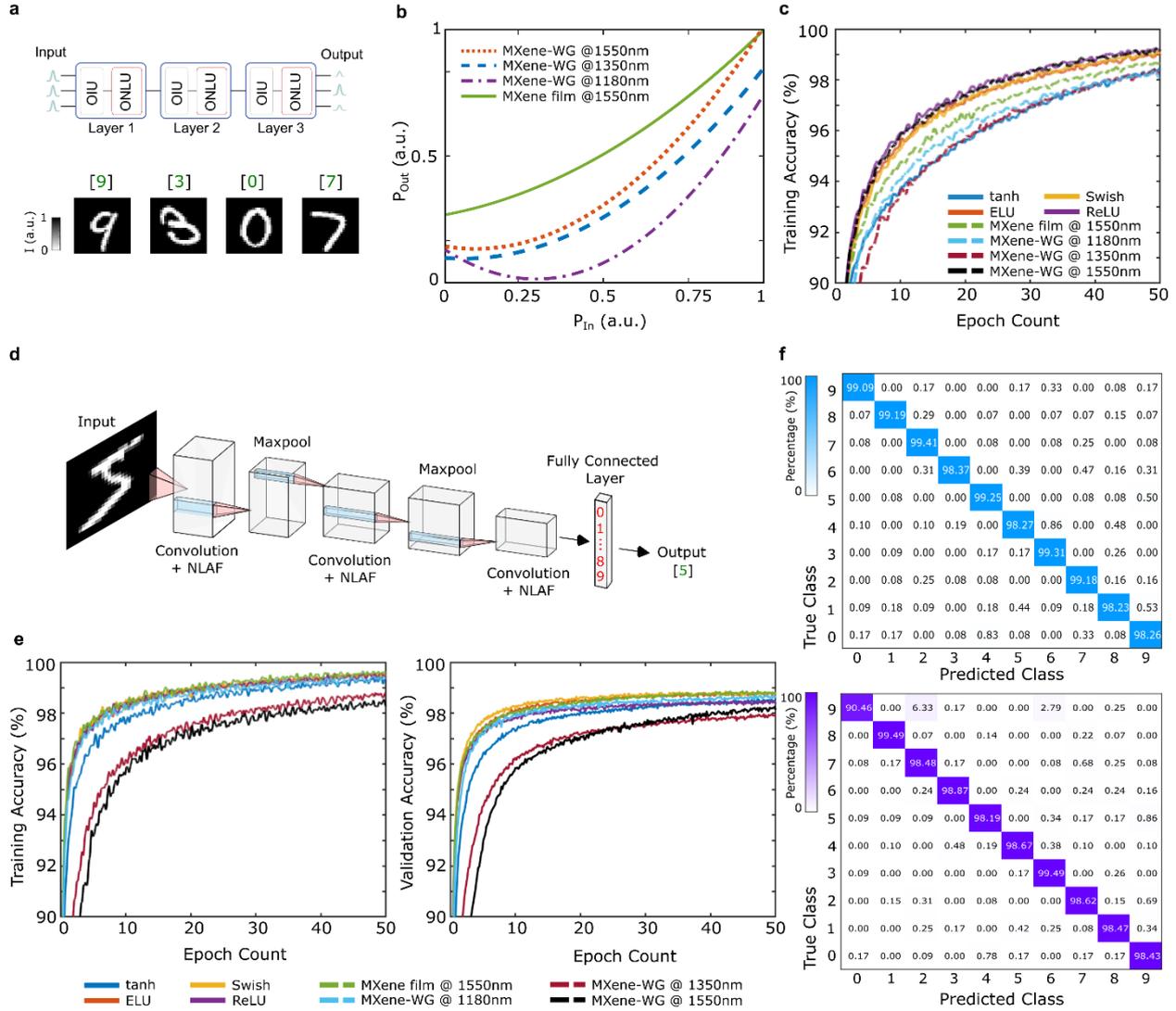

**Figure 5 | High degree of precision MNIST classification with MXene nonlinear activation function. a,** Schematics of the emulated three-layer structure fully connected network. Each layer of the DNN is composed of optical interference and nonlinearity units. (bottom) Several predicted labels correspond to four input handwritten digit images. **b,** Different MXene-based activation functions represent "power in" to "power out" relation for various operating wavelengths. **c,** Network prediction accuracy as a function of epoch count during the training process, with proposed all-optical nonlinear activation functions considering MXene metasurface overlayer on waveguide and MXene thin film compared to standard software-based nonlinear activation functions. **d,** Schematic illustration of decomposes the studied convolutional neural network into unique layers with all-optical nonlinear activation function operations. The input is a handwritten digit image resulting in outputs in the range of 0 to 9. **e,** Model accuracy comparison between the proposed all-optical MXene-based and software-based nonlinear activation functions during the training and validation processes. **f,** Experimental confusion matrices of the MNIST classification, using the transfer functions of MXene thin film (blue) and MXene metasurface overlayer on waveguide (purple) at a wavelength of 1550 nm. The model achieved 98.9% and 97.4% recognition accuracy of the test dataset, respectively.

# Supporting Information for

# $Ti_3C_2T_x$ MXene Enabled All-Optical Nonlinear Activation Function for On-Chip Photonic Deep Neural Networks

*Adir Hazan[1], Barak Ratzker[2], Danzhen Zhang[3], Aviad Katiyi[1], Nachum Frage[2], Maxim Sokol[4], Yury Gogotsi[3] and Alina Karabchevsky[1,*]*

[1] School of Electrical and Computer Engineering, Electro-Optics and Photonics Engineering Department, Ben-Gurion University of the Negev, Beer-Sheva 8410501, Israel.

[2] Materials Engineering Department, Ben-Gurion University of the Negev, Beer-Sheva 8410501, Israel.

[3] Materials Science and Engineering Department, and A.J. Drexel Nanomaterials Institute, Drexel University, Philadelphia 19104, USA.

[4] Materials Science and Engineering Department, Tel Aviv University, Ramat Aviv 6997801, Israel.

∗ Correspondence: alinak@bgu.ac.il

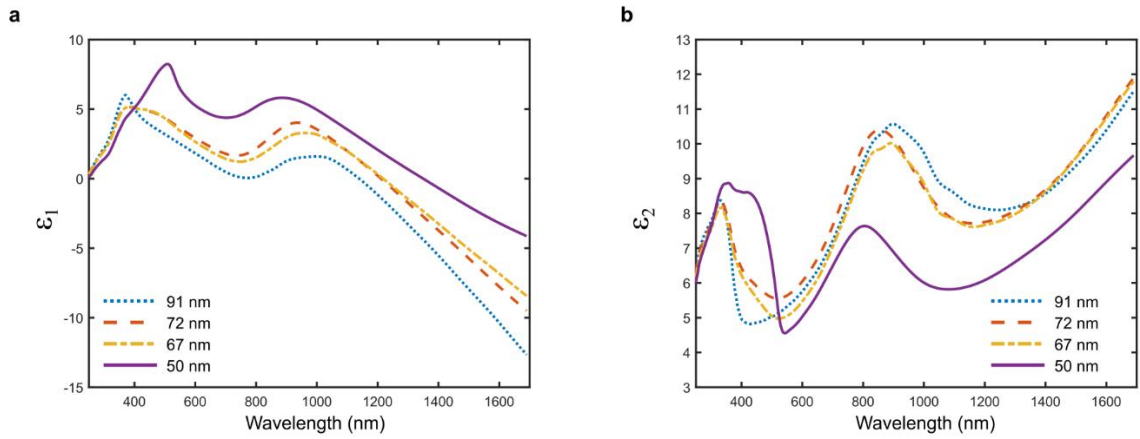

**Supplementary Fig. S1 | Dispersion of MXene thin films. a-b,** The complex permittivity $\tilde{\varepsilon}(\lambda)$, with real $\varepsilon_1$ (**a**) and imaginary $\varepsilon_2$ (**b**) parts as a function of the wavelength for different film thicknesses: 50 nm (solid purple curve), 67 nm (yellow dash-dotted curve), 72 nm (orange dashed curve), and 91 nm (blue dotted curve).

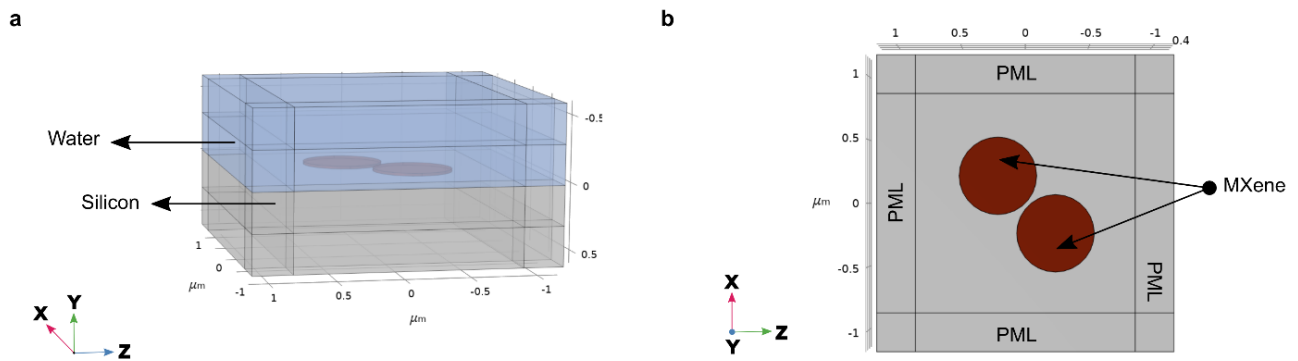

**Supplementary Fig. S2: The nanodisks arrangement. a-b,** A unit cell including two nanodiscs embedded in water (light blue medium) located on top of silicon (grey medium) surface (**a**) as simulated in the numerical model. Considering the light propagating along with the waveguide core (z-direction), with evanescent field components in the y-direction extending into the sample medium. The MXene nanodiscs (red) are separated by 15 nm (**b**). Perfectly matched layers (PML).

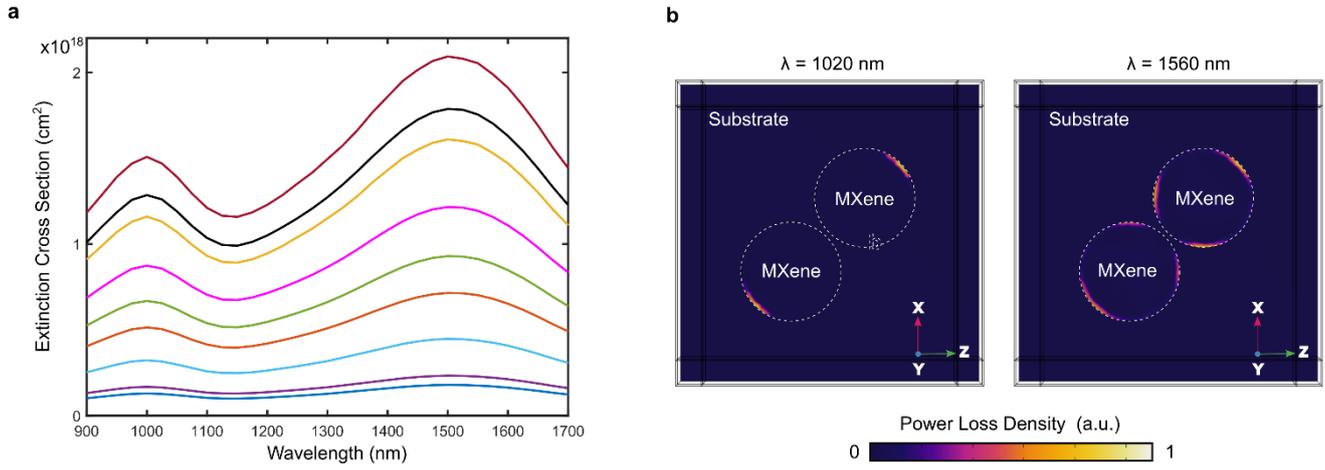

**Supplementary Fig. S3: MXene on a substrate. a,** The model calculated results of extinction cross-section spectra of MXene nanodiscs atop the waveguide for different input powers. **b,** Normalised power loss density in a slice through the MXene nanodiscs (the silicon-water interface) at lower (1020 nm) and longer (1560 nm) wavelengths resonances.

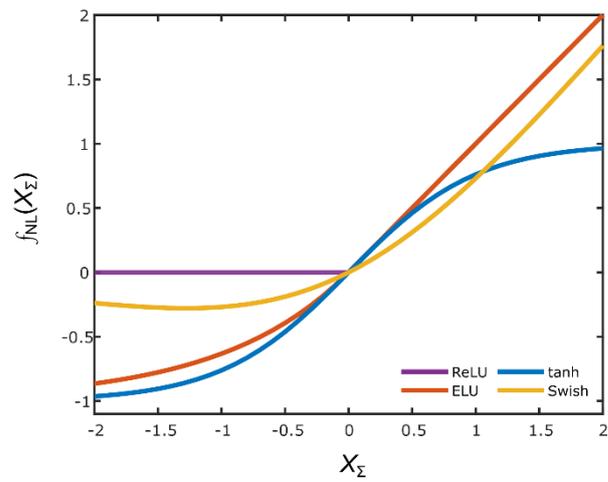

**Supplementary Fig. S4**: **Different common software-based nonlinear activation functions**. Represented by input/output relation: ReLU (purple), ELU (orange), tanh (blue), and Swish (yellow).

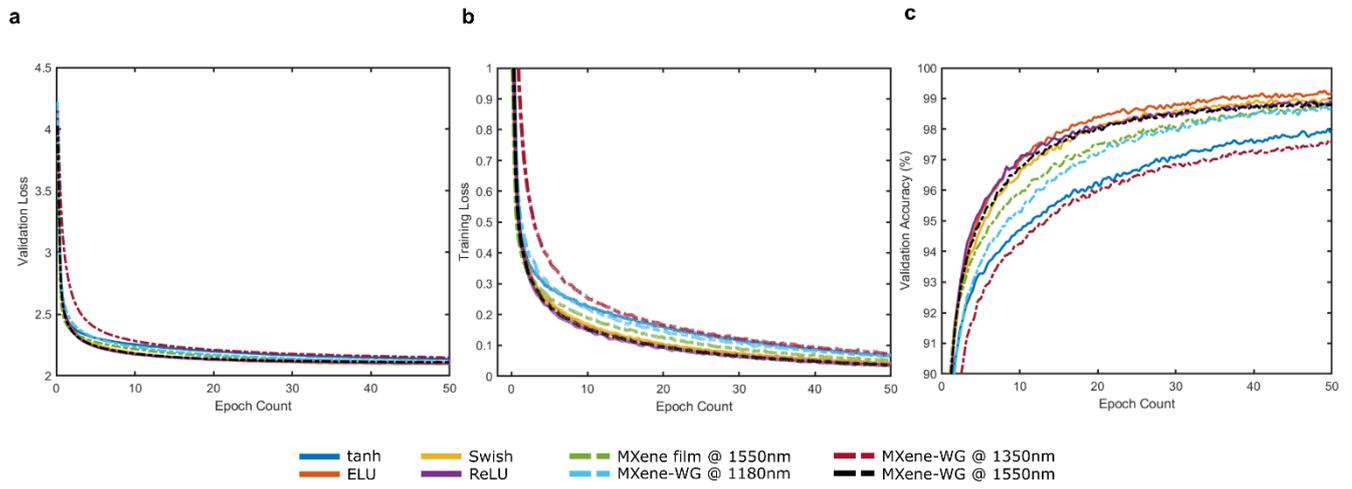

**Supplementary Fig. S5 | Fully connected network for handwritten MNIST digit classification. a-c,** Comparisons of loss as a function of epoch count during the training (**a**) and validation (**b**) processes and model accuracy (**c**), with proposed all-optical nonlinear activation functions considering MXene metasurface overlayer on waveguide and MXene thin film as compared to software-based nonlinear activation functions.

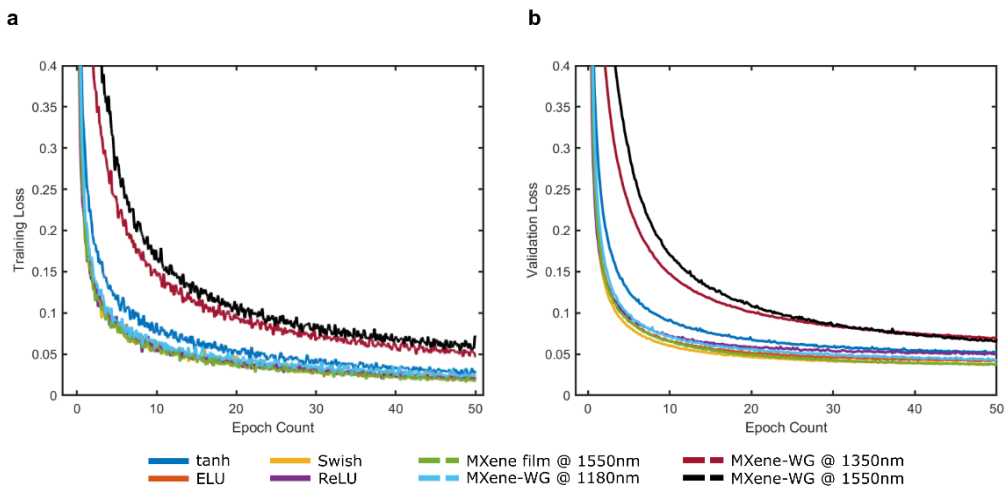

**Supplementary Fig. S6 | CNN network for handwritten MNIST digit classification. a-c,** Comparisons of loss as a function of epoch count during the training (**a**) and validation (**b**) processes, with proposed all-optical nonlinear activation functions considering a MXene metasurface overlayer on waveguide and a MXene thin film as compared to software-based nonlinear activation functions.